\documentclass[a4paper,prc,twocolumn,showpacs,superscriptaddress,floatfix]{revtex4-1}
\usepackage{graphicx}%
\usepackage{dcolumn}%
\usepackage{bm}%
\usepackage{hyperref}%
\hypersetup{
  colorlinks=true,        %
  linkcolor=blue,         %
  citecolor=cyan,         %
  urlcolor=cyan            %
}
\usepackage{colortbl}
\usepackage{amsmath,amssymb,amsfonts}
\usepackage{times}

\newcommand{\dd}{\mathrm{d}}

\begin{document}

\title{A dynamical inflaton coupled to strongly interacting matter}
\author{Christian Ecker}
\affiliation{Institut f\"ur Theoretische Physik, Goethe Universit\"at, Max-von-Laue-Str. 1, 60438 Frankfurt am Main, Germany}
\author{Elias Kiritsis}
\affiliation{Universit\'e Paris Cit\' e, CNRS, Astroparticule et Cosmologie, F-75013 Paris, France}
\affiliation{Institute for Theoretical and Computational Physics, Department of Physics, P.O. Box 2208,University of Crete, 70013, Heraklion, Greece}
\author{Wilke van der Schee}
\affiliation{Theoretical Physics Department, CERN, CH-1211 Gen\`eve 23, Switzerland}
\affiliation{Institute for Theoretical Physics, Utrecht University, 3584 CC Utrecht, The Netherlands}

\date{\today}
 
\begin{abstract}
According to the inflationary theory of cosmology, most elementary particles in the current universe were created during a period of reheating after inflation.
In this work we self-consistently couple the Einstein-inflaton equations to a strongly coupled quantum field theory (QFT) as described by holography. We show that this leads to an inflating universe, a reheating phase and finally a universe dominated by the QFT in thermal equilibrium.
\end{abstract}

\maketitle

\section{Introduction}\label{sec:1}
\vspace{-0.25cm}
Cosmological inflation is a paradigm of extended exponential expansion of our universe at its earliest moments of existence.
Due to the rapid expansion this leads to a quickly cooling universe, which at the end reheats to the hot plasma that then forms the Big Bang. 
One of the main features 
of inflation is that the exponential expansion naturally explains why our universe is to a good approximation homogeneous, even though different parts could not have been in causal contact since the Big Bang.

One of the main uncertainties in inflation is the so-called `exit' to the hot Big Bang scenario.
Due to the many e-foldings of expansion a natural end state of inflation would be an empty universe, so the question is how ordinary and possibly dark matter arise in inflation.
\emph{Standard inflation} posits a distinct reheating stage where the inflaton undergoes a damped oscillation in the inflaton potential while interacting with and heating up ordinary matter \cite{Dolgov:1989us, Traschen:1990sw, Kofman:1994rk, Shtanov:1994ce, Kofman:1997yn} (see \cite{Amin:2014eta, Mazumdar:2018dfl, Bassett:2005xm, Baumann:2009ds, Allahverdi:2010xz} for reviews).
A different scenario is called \emph{warm inflation} \cite{Berera:1995wh, Bastero-Gil:2009sdq, Berghaus:2019whh,Kamali:2023lzq}.
In this case there is always a subdominant but significant part of the universe made up by ordinary or dark matter.
It is only when the inflaton rolls down the potential that subsequently ordinary matter becomes dominant, thereby making a smooth transition to the Big Bang.

Many microscopic models have been proposed for either scenario, all of which have advantages and disadvantages (see e.g. \cite{Mazumdar:2010sa}).
In standard inflation there is often a `preheating' phase, where bosonic fields undergo an exponential increase in density due to resonant amplification.
This, however, leads to a non-thermal state of which it is not \emph{a priori} clear if it thermalizes in time for the Big Bang scenario.
Recently there has been renewed interest in warm inflation, since it may avoid some of the conjectured constraints on consistent quantum gravity theories that arise from the swampland program \cite{Motaharfar:2018zyb, Das:2019acf}.

In this work we present a toy universe in which the inflaton is coupled to a strongly coupled QFT (see also \cite{Kawai:2015lja} for an earlier attempt).
A unique and important aspect of strongly coupled QFTs is that they approach hydrodynamics and thermalize as fast as possible~\cite{Heller:2011ju, Liu:2018crr}.
At the relevant energy scales even the strong coupling constant of Quantum Chromodynamics is small due to asymptotic freedom, so this QFT can be thought of as a hidden sector that exists at
some high energy scale.
The strongly coupled QFT is described using holography, which is a remarkable duality arising from string theory that can describe strongly coupled QFTs in terms of a classical anti-de-Sitter (AdS) universe of one higher dimension.
The extra dimension can be thought of as energy scale, whereby for a thermal state there exists a black hole horizon in the infrared.

\begin{figure}[htb]
 \center
 \includegraphics[width=.95\linewidth]{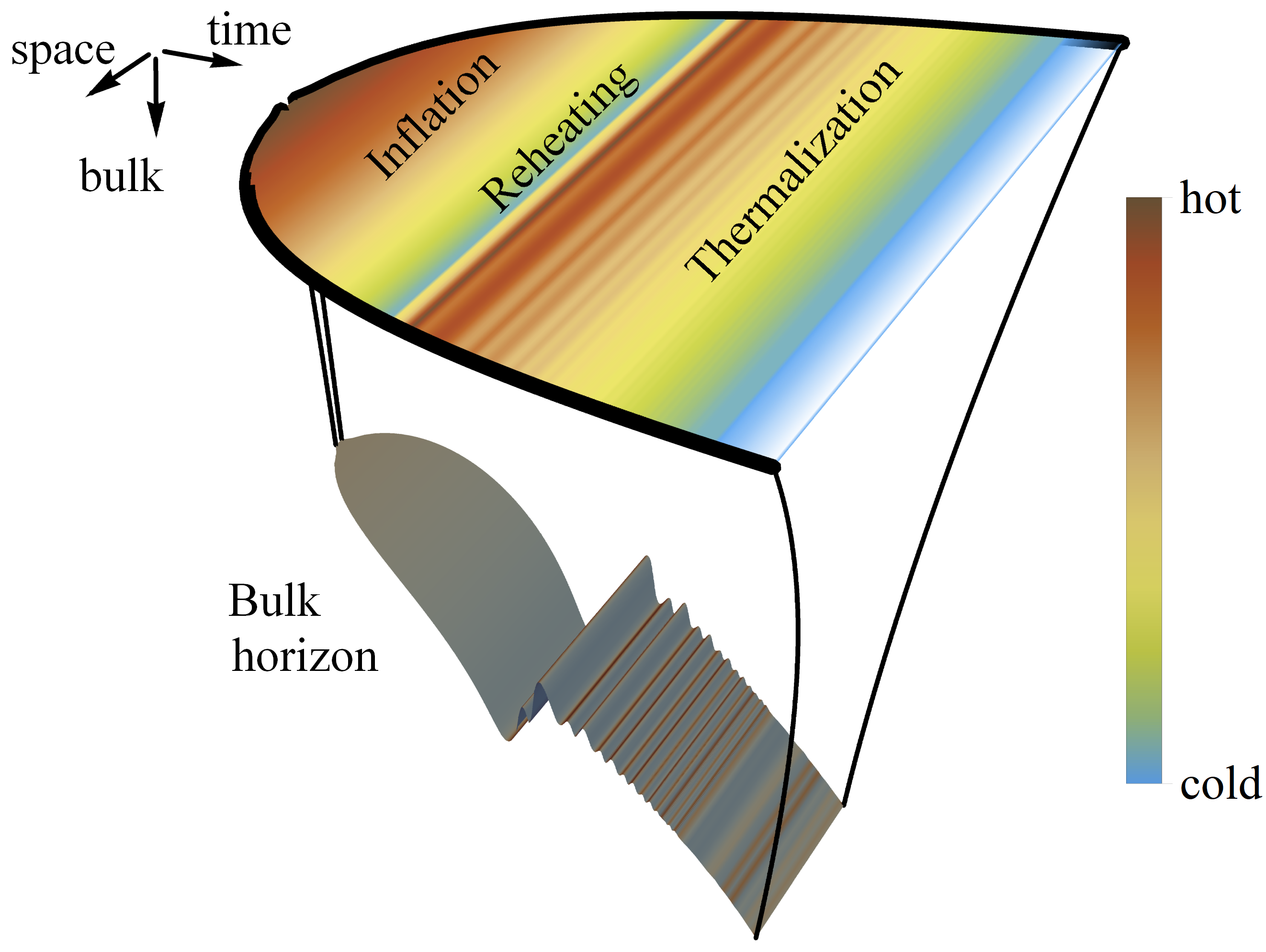}
 \caption{Illustration of the bulk and boundary during holographic reheating.}
 \label{fig:Illustr}
\end{figure}

While we present a general framework for reheating with a strongly coupled QFT, in this work we will present a simple model example to illustrate its dynamics.
Quite strikingly we find that the model qualitatively reproduces many of the features of warm inflation (see Fig.~\ref{fig:Illustr} for a cartoon).
This includes an extended period of cooling and exponential expansion, an inflaton rolling down the potential, heating up the QFT and finally the transition to a universe dominated by QFT matter in a thermal state.

While in this Letter we use standard inflationary terminology in describing the evolution of the constructed universe we stress that in this work we do not attempt to construct a realistic model for our universe. 
Rather, we focus on a qualitative general description of an inflaton interacting with a strongly coupled QFT with a specific evolution as an explicit example.

\section{Model}\label{sec:2}
In order to model the energy transfer of the inflaton field to matter on a dynamical spacetime, we evolve self-consistently the Einstein-inflaton equations together with the energy momentum tensor for strongly coupled matter described by holography.   
The total action of this model consists of two different sectors and an interaction part
\begin{equation}\label{eq:Stot}
 S=S_{\rm EH + inf}+S_{\rm hol}+S_{\rm int}\,.
\end{equation}
The first sector $S_{\rm EH + inf}$ consists of four-dimenensional Einstein gravity with a dynamical inflaton field, $S_{\rm hol}$ models the dynamics of a strongly coupled QFT via the gauge/gravity duality in terms of a five-dimensional gravity dual and $S_{\rm int}$ accounts for the direct coupling between these two sectors.

\begin{figure}[htb]
 \center
 \includegraphics[width=0.8\linewidth]{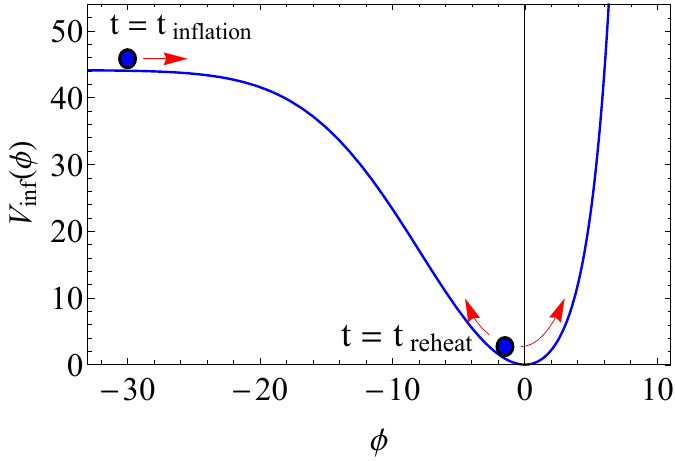}
 \caption{We show the inflaton potential. As an initial condition the inflaton starts at $\phi = -30$ and then slowly rolls down the potential. At the bottom of the potential the inflaton oscillates and reheats the universe due to its coupling to the QFT.}
 \label{fig:Vinflaton}
\end{figure}

The first term in~\eqref{eq:Stot} is the standard Einstein-Hilbert plus Klein-Gordon action with a non-trivial scalar field potential $V_{\rm inf}(\phi)$ which together describe the dynamics of the spacetime and the inflaton $\phi$ in the four boundary dimensions
\begin{equation}
  S_{\rm EH + inf}=\int\dd^4x\sqrt{-\gamma}\left(\frac{R}{2\kappa_4} -\frac{1}{2}\gamma^{ij}\partial_i\phi\partial_j\phi-V_{\rm inf}(\phi)\right)\,.
\end{equation}
Here $\kappa_4$ parametrizes the strength of the gravitational interaction and $R$ is the Ricci scalar of the spacetime metric $\gamma_{ij}$.
We define this metric to be of Friedmann--Lemaître--Robertson--Walker (FLRW) type
\begin{equation}\label{eq:FLRW}
 \dd s^2=\gamma_{ij}\dd x^i \dd x^j=- \dd t^2 + a(t)^2 \dd\vec{x}^2\,,
\end{equation}
where $a(t)$ is the scale factor that determines the expansion of the spacetime via the Einstein field equations.
We consider a generic family of inflaton potentials given by
\begin{equation}
 V_{\rm inf}(\phi)=v_0 + v_1\,e^{v_2(\phi-\phi_m)}-v_3\,e^{-v_4 (\phi-\phi_m)^2}\,,
\end{equation}
where $\phi_m$ and %
$v_0$ are fixed by demanding that the potential and the inflaton vanish at the global minimum $V_{\rm inf}(0)=V_{\rm inf}'(0)=0$.
In Fig.~\ref{fig:Vinflaton} we show the potential we use in our simulation, where we set
$v_1=9/8$, $v_2=2/3$, $v_3=45$ and $v_4=1/50$. 
This potential allows an inflaton that starts at $\phi=-30$ to go through a sufficiently long inflating phase followed by an oscillating phase in which it then reheats the QFT.

The strongly coupled matter sector is defined via the gauge/gravity duality by the five-dimensional bulk action
\begin{equation}\label{eq:Sbulk}
 S_{\rm bulk}=\frac{2}{\kappa_5}\int \dd^5x\sqrt{-g}\left(\frac{1}{4}\mathcal{R}-\frac{1}{2}(\partial\Phi)^2-V_{\rm bulk}(\Phi)\right)\,,
\end{equation}
where $\kappa_5$ denotes the bulk gravitational coupling, $\mathcal{R}$ is the Ricci scalar associated to the bulk metric $g_{\mu\nu}$ and $\Phi$ is a bulk scalar field with potential 
\begin{equation}\label{eq:defW}
V_{\rm bulk}(\Phi)=\frac{1}{L^2}\left(-3-\frac{3 \Phi ^2}{2}-\frac{\Phi ^4}{3}+\frac{11 \Phi ^6}{96}-\frac{\Phi ^8}{192}\right)\,,
 \end{equation}
 where $L$ denotes the length scale of the asymptotic AdS metric $g_{\mu\nu}$, which we set to unity. 
The bare bulk action $S_{\rm bulk}$ needs to be renomalized by adding appropriate counter terms $S_{\rm bdry}$ which render the holographic action $S_{\rm hol}=S_{\rm bulk}+S_{\rm bdry}$ in Eq.~\eqref{eq:Stot} finite on-shell.
The renormalized action $S_{\rm hol}$ then defines a holographic bottom-up model~\cite{Attems:2016ugt} for a strongly coupled QFT with regular renormalization group flow between its conformal ultraviolet and infrared fixed points and broken conformal symmetry in between.
The mass of the bulk scalar field $m^2=\frac{\partial^2 V}{\partial \Phi^2}|_{\Phi=0}=-\frac{3}{2}$ determines the conformal scaling dimension $\Delta=3$ of the dual operator $\mathcal{O}$ via the relation $m=\sqrt{\Delta(\Delta-4)}$.

\begin{figure*}[htb]
\center
\includegraphics[width=0.325\linewidth]{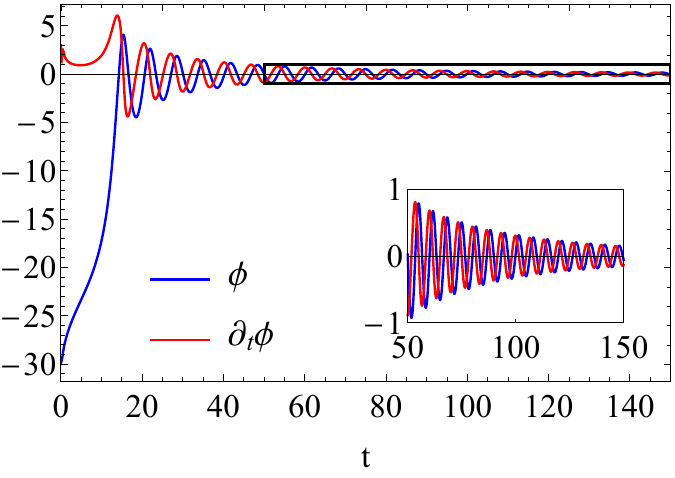}
\includegraphics[width=0.325\linewidth]{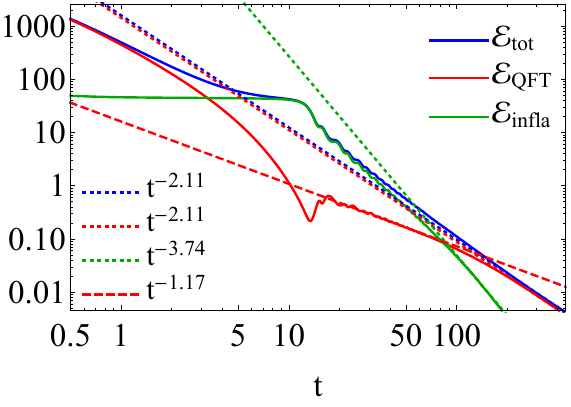}
\includegraphics[width=0.325\linewidth]{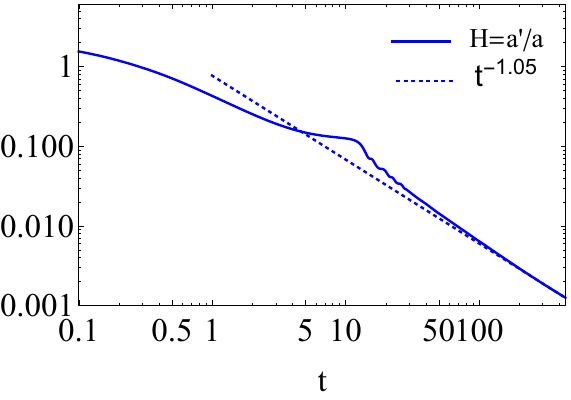}
\caption{(left) After an initial stage where the QFT cools down (till about $t=3$), the inflaton $\phi$ slow-rolls down till it starts oscillating in the potential well.
(middle) Initially the dynamics is dominated by the QFT energy till about $t = 3$.
After this the universe inflates till the inflaton reaches the bottom of the potential at $t = 14.3$.
The inflaton oscillations then reheat the QFT universe. 
(right) Initially the Hubble rate decreases due to the dilution of the QFT energy till $t\approx 5$. After this the universe inflates at a constant exponential rate till about $t = 14.3$ when the inflaton is at the bottom of the potential.
}
\label{fig:energy}
\end{figure*}

Finally, there is an interaction term in the effective action that couples the inflaton via the vacuum expectation value (VEV) of the scalar operator $\langle \mathcal{O}\rangle=\mathcal{O}_{\rm QFT}/\kappa_5$ to the holographic sector
\begin{equation}\label{eq:Sint}
 S_{\rm int}=\int\dd^4x\sqrt{-\gamma}~U(\phi)~\mathcal{O}_{\rm QFT}\,,
\end{equation}
where the free function $U(\phi)$ defines the coupling of the model. 
In the QFT the inflaton hence acts as a source for the scalar operator $\mathcal{O}$, where the source is given by the asymptotic boundary value of the bulk scalar $\Phi_{(0)}=U(\phi)$.

The total energy momentum tensor in the boundary theory consists of three parts
\begin{equation}
\label{eq:fullTmunu}
 T_{ij}={\rm diag}\left(\mathcal{E},\mathcal{P},\mathcal{P},\mathcal{P}\right)=T^{\rm inf}_{ij}+\mathcal{T}_{ij}^{\rm QFT}+T^{\rm int}_{ij}\,,
\end{equation}
where $\mathcal{E}$ and $\mathcal{P}$ denote energy density and pressure, respectively. 
The first part is the usual expression for the energy momentum tensor of a scalar field
\begin{equation}
 T^{\rm inf}_{ij}=\partial_i\phi\partial_j\phi-\gamma_{ij}\left(\frac{1}{2}\partial_k\phi\partial^k\phi+V_{\rm inf}\right)\,.
\end{equation}
The second term $\mathcal{T}_{ij}^{\rm QFT}={\rm diag}\left(\mathcal{E}_{\rm QFT},\mathcal{P}_{\rm QFT},\mathcal{P}_{\rm QFT},\mathcal{P}_{\rm QFT}\right)$ is the VEV of the holographic energy momentum tensor, where $\mathcal{E}_{\rm QFT}$ and $\mathcal{P}_{\rm QFT}$ denote the corresponding energy density and pressure.
The third term results in an energy-momentum contribution due to the direct coupling between the inflaton and the holographic sector
\begin{equation}
 T^{\rm int}_{ij}=-\frac{2}{\sqrt{-\gamma}}\frac{\delta S_{\rm int}}{\delta \gamma^{ij}}=U(\phi)\,\mathcal{O}_{\rm QFT}\,\gamma_{ij}\,. 
\end{equation}
The total energy momentum tensor is covariantly conserved
\begin{equation}
\nabla^iT_{ij}=\nabla^i T^{\rm inf}_{ij} +\nabla^i \mathcal{T}_{ij}^{\rm QFT}+\nabla^i T^{\rm int}_{ij}=0\,,
\end{equation}
when using the on-shell condition for the scalar field $\nabla^i T^{\rm inf}_{ij}=-U(\phi) \,\partial_j \mathcal{O}_{\rm QFT}$ together with the Ward identity for the holographic energy momentum tensor $\nabla^i\mathcal{T}_{ij}^{\rm QFT}=-\partial_j\,U(\phi)\,\mathcal{O}_{\rm QFT}$, where $\nabla^i$ is the Levi-Civita connection associated to $\gamma_{ij}$.

In the standard holographic dictionary the QFT lives on a fixed (curved) background spacetime $\gamma_{ij}$ and also the scalar source $\Phi_{(0)}$ acts as a free parameter that can be specified arbitrarily. 
Here, however, we require them to satisfy the equations of motion that follow from the boundary action \eqref{eq:Stot}.
For the FLRW line element~\eqref{eq:FLRW} one obtains the standard Friedmann equations together with a scalar field equation for the inflaton that is coupled to $\mathcal{O}_{\rm QFT}$:
\begin{align}
  H(t)^2 & =  -\frac{\kappa_4}{3} \mathcal{E}(t)\,,\label{eq:Feq1} \\
 \frac{a''(t)}{a(t)} & =  -\frac{1}{2}\left(\kappa_4 \mathcal{P}(t)+H(t)^2\right)\,,\label{eq:Feq2} \\
 \phi ''(t) & = \partial_\phi\,U(\phi(t))\,\mathcal{O}_{\rm QFT}(t) -3 H(t) \phi '(t)-\partial_\phi V_{\rm inf}(\phi (t))\,,\label{eq:Feq3}
\end{align}
where $H(t)=a'(t)/a(t)$ is the Hubble rate and the total energy density and pressure are given by
\begin{eqnarray}
\mathcal{E}&=&\mathcal{E}_{\rm QFT} + V_{\rm inf}(\phi) + U(\phi)\,\mathcal{O}_{\rm QFT} + \frac{1}{2} \phi'^2\,,\\
\mathcal{P}&=&\mathcal{P}_{\rm QFT} - V_{\rm inf}(\phi) - U(\phi)\,\mathcal{O}_{\rm QFT} + \frac{1}{2} \phi'^2\,.
\end{eqnarray} 
The $-3\,H(t)\,\phi '(t)$ in Eq.~\eqref{eq:Feq3} is the standard friction term that brings the inflaton to rest, but we note that with the holographic coupling the scalar VEV $\mathcal{O}_{\rm QFT}$ also contributes. 
Importantly, both $\mathcal{E}_{\rm QFT}$ and $\mathcal{O}_{\rm QFT}$  depend on the full bulk geometry, including explicit dependencies on $\phi(t)$, $\phi '(t)$, $a(t)$ and $a'(t)$. 
In addition, $\mathcal{O}_{\rm QFT}$, $\mathcal{E}_{\rm QFT}$ and $\mathcal{P}_{\rm QFT}$ are not independent, but related via the trace Ward identity
\begin{equation}
\gamma^{ij}\mathcal{T}_{ij}^{\rm QFT}=\mathcal{E}_{\rm QFT}-3\,\mathcal{P}_{\rm QFT}=-U(\phi)\,\mathcal{O}_{\rm QFT}+\mathcal{A}\,,
\end{equation}
where $\mathcal{A}$ is the conformal anomaly~\cite{Henningson:1998gx,Papadimitriou:2011qb}.
The variational principle in holography with dynamical boundary conditions, the holographic renormalization of $S_{\rm hol}$, together with the resulting expressions for $\mathcal{E}_{\rm QFT}$, $\mathcal{P}_{\rm QFT}$, $\mathcal{O}_{\rm QFT}$ and  the corresponding anomaly corrected Ward identities
as well as the thermodynamic properties of the holographic QFT 
are given in the Supplemental Material.

\begin{figure*}[htb]
\center
\includegraphics[width=0.94\linewidth]{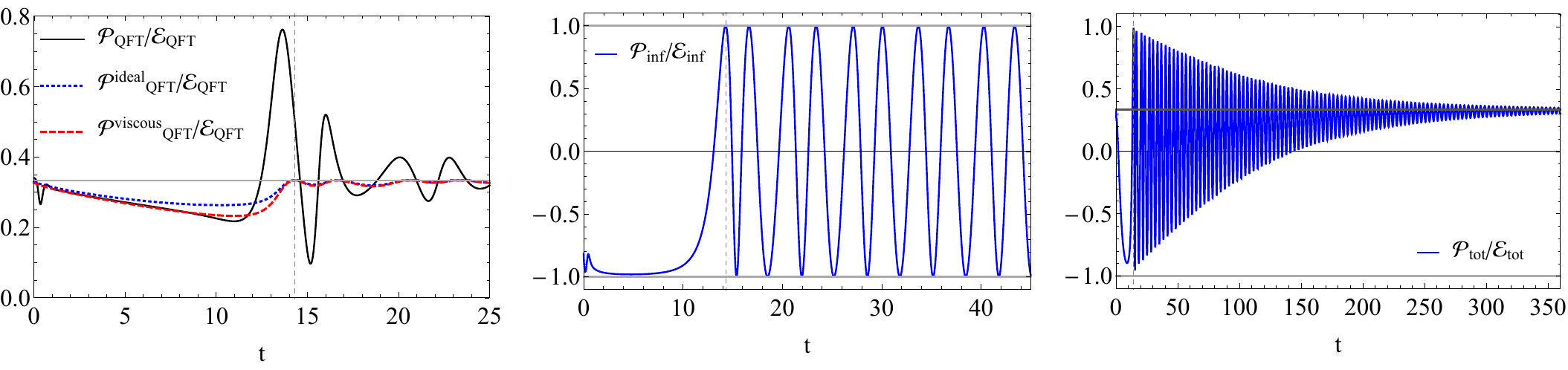}
\caption{(left) The pressure over energy density of the QFT together with the predictions from ideal and viscous hydrodynamics. 
After a brief initial hydrodynamisation period, the QFT is well described by viscous hydrodynamics until the inflaton sources the QFT out of equilibrium.
(middle) The equivalent figure for the inflaton.
Initially it is dominated by the potential having $\mathcal{P}=-\mathcal{E}$ after which it oscillates around the minimim.
(right) We show the total pressure over energy density, which is initially dominated by the QFT, then by the inflaton and at late times again by the reheated QFT.
}
\label{fig:pressures}
\end{figure*}

\begin{figure}[htb]
\center
\includegraphics[width=0.75\linewidth]{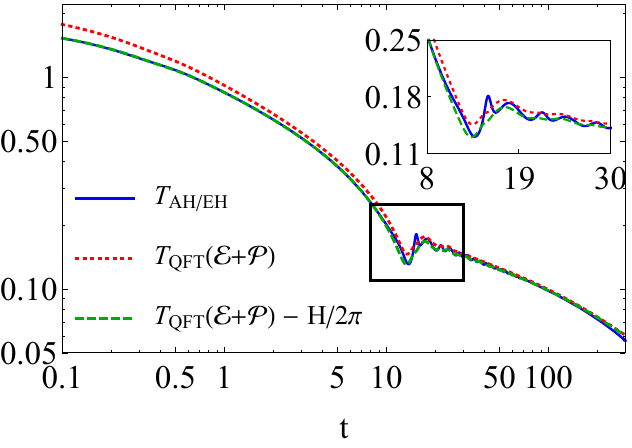}
\caption{We show in blue the temperature as measured by the surface gravity of the apparent horizon (AH) and the event horizon (EH), which are numerically indistinguishable. The dotted red line is the temperature of the QFT as determined by the equation of state. 
The green dashed line shows that the horizon temperatures are lower than the QFT temperature by exactly the de Sitter temperature $T_{\rm ds} = H/2\pi$.
}
\label{fig:temperature}
\end{figure}

\section{Solution Method}\label{sec:2.3}
Computing the time evolution of the scale factor $a(t)$, the inflaton $\phi(t)$ and the energy-momentum tensor $T_{ij}(t)$ for a given set of initial conditions, requires to solve the corresponding initial value problem for Eqs.~\eqref{eq:Feq1} to \eqref{eq:Feq3} together with the dual bulk initial-boundary value problem in a self-consistent way.
For this we follow essentially the same procedure as in \cite{Ecker:2021cvz}, where we solve a similar system with dynamical boundary metric, but with constant source $\Phi_{(0)}$ for the scalar field operator.
The only addition here is that we promote $\Phi_{(0)}(t)=U(\phi(t))$ to a dynamical field whose time dependence is determined by the inflaton equation of motion~\eqref{eq:Feq3}. 

At the initial time $t=t_{\rm ini}$ we need to specify initial conditions for the energy density $\mathcal{E}_{\rm QFT}(t_{\rm ini})=\mathcal{E}_{\rm QFT}^{\rm ini}$, the inflaton $\phi(t_{\rm ini})=\phi_{\rm ini}$ and its time derivative $\partial_t\phi(t_{\rm ini})=\phi'_{\rm ini}$ as well as a profile for the bulk scalar $\Phi(r,t_{\rm ini})=\Phi_{\rm ini}(r)$ along the holographic coordinate $r$ and whose asymptotic value is consistent with the inflaton $\lim_{r\to\infty}\,r\,\Phi_{\rm ini}(r)=U(\phi_{\rm ini})$.
Eqs.~\eqref{eq:Feq1} to \eqref{eq:Feq3} then determine $\phi''_{\rm ini}$ as well as the Hubble rate $H$ and its time derivative $H'$.
It is important to note that $\mathcal{O}_{\rm QFT}$ depends on $H'$ and also that $\mathcal{E}_{\rm QFT}$ depends on $\partial_t^2\phi$. 
The equations are hence coupled and lead to a sixth order polynomial equation which we solve numerically~\footnote{We use \emph{Mathematica} to numerically solve for the six different solutions for the Hubble rate and take the negative real solution with smallest absolute magnitude to construct the initial data.}.
After the initialisation we evolve $H$ and $\partial_t \phi$ using Eqs.~\eqref{eq:Feq2} and \eqref{eq:Feq3}.
As in \cite{Ecker:2021cvz} for the boundary metric we replace $\partial_t^3 \phi$ and $\partial_t^4 \phi$ derivatives that appear in the regularized bulk equations by their solutions in terms of the near-boundary expansion.

For the evolution presented in this work we set $\kappa_5=1/9$, $\kappa_4=\frac{2\pi}{5625}$, $U(\phi)=\lambda\,\phi$ with $\lambda = 1/30$ and use $\mathcal{E}_{\rm QFT}^{\rm ini}=13275$, $\phi_{\rm ini}=-30$, $\phi'_{\rm ini}=3/10$ and $\tilde{\Phi}(r)=-6+120/r-300/r^3$ as initial conditions, where $\tilde{\Phi}(r)$ is defined by $\Phi(r) \equiv \Phi_{NB}(r)+r^{-3}\tilde{\Phi}(r)$ and $\Phi_{NB}(r)$ contains near-boundary terms up to $\mathcal{O}(r^{-2})$ and $\mathcal{O}(r^{-4} \log(r))$. These parameters are tuned to get an evolution that shows both an inflationary and a reheating phase.
Key choices for the evolution to be numerically stable are to start with a relatively high energy density, which guarantees a large (and hence stable) bulk black hole horizon for a relatively long time. 
The energy density should however not be so high that it would dominate the inflaton dynamics for the entire evolution.
In this way the energy density at the end of the inflationary period is close to the vacuum and significant effects of the inflaton coupling can be seen to heat up the QFT in the reheating phase.

\section{Results}\label{sec:3}
Fig.~\ref{fig:energy} shows the resulting evolution of the inflaton (left), energy density (middle) and Hubble rate (right) of the model.
The early phase is dominated by the high initial energy density of the QFT, but at $t=3.27$ the inflaton energy density becomes dominant and the universe enters a phase of relatively constant exponential expansion.
Later at $t=14.3$ the inflaton reaches the bottom of the potential and starts oscillating rapidly.
These oscillations form sources for the QFT energy, which then increases from a minimum of $\mathcal{E}_{\rm QFT}=0.21$ at $t=13.5$ to a subsequent maximum of $\mathcal{E}_{\rm QFT}=0.64$ at $t=17.3$.
Crucially this reheating continues, which is apparent from the relatively slow scaling $\mathcal{E}_{\rm QFT}\propto t^{-1.17}$ of the QFT energy density.
The universe then keeps expanding at increasingly slower rates, thereby cooling down both the inflaton and the QFT energy density.
At late times the QFT energy density is dominant.

In Fig.~\ref{fig:pressures} we show the evolution of the pressure of the QFT (left), the inflaton (middle) as well as the total pressure (right).
After a very short far-from-equilibrium stage, we see that the QFT pressure is well described by the equations as given by viscous hydrodynamics, much like what was found in \cite{Ecker:2021cvz}.
After the inflaton rolls down, however, we see that the reheating pushes the QFT significantly out of equilibrium.
After this the system settles down to equilibrium rather quickly.
At late times the evolution is completely dominated by the QFT, which is now close to its conformal IR fixed point where $\mathcal{P_{\rm QFT}}=\mathcal{E_{\rm QFT}}/3$ (Fig.~\ref{fig:pressures} right).

In Fig.~\ref{fig:temperature} we show in blue the temperature obtained from the surface gravity of the apparent horizon $T_{\rm AH}$ (explicit formulas are given in  the Supplemental Material). 
We verified that the event horizon location is numerically indistinguishable from the apparent horizon throughout the evolution, which is expected for a theory in thermal equilibrium but unlike the vacuum de Sitter case of \cite{Casalderrey-Solana:2020vls}.
During the entire evolution the temperature is dominated by the QFT energy $\mathcal{E}$.
Since $H^2 \propto \mathcal{E}$ and $T_{\rm QFT}^4 \propto \mathcal{E}$ at late times, the temperature of the cosmological horizon $T_{\rm dS} = H/2\pi \propto T_{\rm QFT}^2$ is negligible if $T_{\rm QFT}$ is small.
At early times we notice a significant difference between the apparent horizon temperature and the temperature obtained from the QFT equation of state.
This can be fully explained by the fact that the universe is expanding.
Indeed, subtracting $T_{\rm QFT} \rightarrow T_{\rm QFT} - H/2\pi$ accurately describes the complete evolution with the exception of a small off-equilibrium time-window where the inflaton approaches the minimum of the potential for the first time.
This is consistent with the analytical solution of a thermal plasma expanding in de Sitter space for a strongly coupled conformal theory (see e.g. \cite{Apostolopoulos:2008ru, Buchel:2016cbj}).
We verified that the exact same subtraction describes the evolution in Fig.~9 of \cite{Casalderrey-Solana:2020vls} up to the point where $T_{\rm QFT} \approx T_{\rm dS}$.

\section{Discussion}\label{sec:4}
For simplicity, our work is restricted to a specific model which assumes a holographic potential that realizes in the dual field theory a renormalization group flow between UV- and IR-fixed points and leads to the thermodynamics of a smooth cross over between two conformally symmetric phases.

Changing the potential would allow to study QFTs with different equilibrium properties, like for example theories with phase transitions and confinement~\cite{Gursoy:2007er,Gursoy:2008za}, or
one may vary the dimension $\Delta$ of the scalar operator that couples to the inflaton.
Choosing $\Delta<3$ for example, makes the linear coupling to the inflaton relevant, as $\phi$ has a weak-coupling dimension near $1$.

It would also be interesting to generalize the field content of our construction, for example by adding the dynamics of a gauge field in the bulk theory, which would allow to model the dynamics of conserved charges~\cite{Folkestad:2019lam} and gauge fields~\cite{Ecker:2018ucc,Ahn:2022azl} in the boundary theory.

One may also change the function $U(\phi)$ that controls the coupling of the inflaton to the scalar QFT operator.
Non-linear options for this function may affect non-trvially the evolution, like for example a quadratic $U$ affects the effective mass of the inflaton and may stop inflation if it becomes large enough.

The most exciting avenue will be to make our model into a realistic inflationary scenario for our own universe that satisfies all the constraints known from cosmology. 
For this several steps are required, including a realistic coupling of the QFT to fields of the Standard Model.
With the current numerical code it would be challenging to obtain the number of e-foldings that are realistically required, but from Fig.~\ref{fig:pressures} it can be seen that accurate approximations using hydrodynamics may be feasible.

\medskip
\begin{acknowledgments}
We thank Alex Buchel, Valerie Domcke, David Mateos, Francesco Nitti  and Tomislav Prokopec for interesting discussions. C.~E. acknowledges support by the Deutsche
Forschungsgemeinschaft (DFG, German Research Foundation) through the
CRC-TR 211 ``Strong-interaction matter under extreme conditions''--
project number 315477589 -- TRR 211.
E. K. was supported in part by CNRS contract IEA 199430.
\end{acknowledgments}

\bibliography{main}

\section*{Supplemental Material}

\subsection{Variational Principle with Mixed Boundary Conditions}\label{app:VP}
In this appendix we review the variational principle in holography with dynamical boundary conditions for the source fields~\cite{Compere:2008us,Ishibashi:2023luz,Ahn:2022azl}.
We will restrict to the case relevant to this work, namely Einstein-dilaton gravity with dynamical boundary conditions for the metric and the scalar field governed by four-dimensional Einstein-inflaton equations of motion.

In the supergravity limit, the gauge/gravity duality allows to express the partition function of a strongly coupled quantum field theory (QFT) as a path integral of a higher dimensional gravity action
\begin{equation}
Z_{\rm QFT}[\gamma,\phi]=\int[\mathcal{D}g]_\gamma[\mathcal{D}\Phi]_\phi e^{-S_{\rm hol}}\,,
\end{equation}
where $\gamma_{ij}$ and $\phi$ denote the background geometry of the dual QFT and the inflaton field on the boundary, while $\int[\mathcal{D}g]_\gamma[\mathcal{D}\Phi]_\phi$ means integration over all bulk geometries and scalar fields with fixed boundary conditions $\gamma_{ij}$ and $\phi$, respectively.
Promoting $\gamma_{ij}$ and $\phi$ to dynamical fields allows one to define an induced gravity partition function as a path integral over all boundary fields
\begin{align}\label{eq:Zind}
Z_{\rm ind}=&\int \mathcal{D}\gamma\mathcal{D}\phi Z_{\rm QFT}[\gamma,\phi]\nonumber\\
 =&\int \mathcal{D}\gamma\mathcal{D}\phi\int[\mathcal{D}g]_\gamma[\mathcal{D}\Phi]_\phi e^{-S_{\rm hol}}\nonumber\\
 =&\int \mathcal{D}g\mathcal{D}\Phi e^{-S_{\rm hol}}\,.
\end{align}
Because the boundary fields are dynamical, the variations of the action result, in addition to the bulk equations of motion, also in some non-trivial boundary contributions
\begin{eqnarray}
\delta_g S_{\rm hol}&=&\int_{\mathcal{M}}\dd x^5 \sqrt{-g}
\,{\rm EOM_{bulk}^{(g)}}\,\delta g^{\mu\nu}\nonumber\\
&+&\int_{\partial\mathcal{M}}\dd x^4 \sqrt{-\gamma}\frac{1}{2}\langle T_{ij}^{\rm QFT}\rangle\delta\gamma^{ij}\,,\label{eq:dSg}\\
\delta_\Phi S_{\rm hol}&=&\int\dd x^5 \sqrt{-g}
\,{\rm EOM_{bulk}^{(\Phi)}}\,\delta \Phi^{\mu\nu}\nonumber\\
&+&\int_{\partial\mathcal{M}}\dd x^4 \sqrt{-\gamma}\langle \mathcal{O}\rangle\delta\phi\,.\label{eq:dSphi}
\end{eqnarray}
Since $S_{\rm hol}$ includes the counter terms, whose explicit form is derived in the next section, the boundary contributions can be identified with the expectation values of the renormalized holographic stress tensor $\langle T_{ij}^{\rm QFT}\rangle$ and the scalar field operator $\langle\mathcal{O}\rangle$ of the boundary theory.

There are different ways to make the actions \eqref{eq:dSg} and \eqref{eq:dSphi} stationary~\cite{Compere:2008us}.
The simplest option is to impose Dirichlet boundary conditions on $g_{\mu\nu}$ and $\Phi$, i.e., demanding $\delta \gamma_{ij}=0$ and $\delta \phi=0$, which makes the boundary geometry and the inflaton field static. 
Another option is to impose Neuman boundary conditions, which is demanding  $\langle T_{ij}^{\rm QFT}\rangle=\langle \mathcal{O}\rangle=0$. 
In this case $\gamma_{ij}$ and $\phi$ can remain dynamical and the bulk geometry is fixed to empty AdS$_5$.
Combinations of these two options are of course also possible.
Finally, the most general possibility is to impose mixed boundary conditions, that is to demand $\frac{1}{2}\langle T_{ij}^{\rm QFT}\rangle+\frac{\delta S_{{\rm bdry,}\gamma}}{\delta \gamma^{ij}}=0$ and $\langle \mathcal{O}\rangle+\frac{\delta S_{{\rm bdry,}\phi}}{\delta \phi}=0$ for some functionals of the boundary metric $S_{{\rm bdry,}\gamma}$ and the scalar field $S_{{\rm bdry,}\phi}$ that can be added to the bulk action.
In this work we choose these boundary functionals to be given by the Einstein-Hilbert plus inflaton action $S_{\rm EH+inf}[\phi,\gamma_{ij}]$ and the interaction term $S_{\rm int}[\phi,\gamma_{ij}]$.

\subsection{Holographic Renormalization}\label{app:holren}
In this appendix we derive the explicit form of the renormalized expectation values of the holographic energy momentum tensor and the scalar field operator~\cite{Bianchi:2001de,Bianchi:2001kw,Skenderis:2002wp}. 
In the following we assume Fefferman--Graham (FG) gauge
\begin{equation}\label{metricFG}
ds^2= L^2\frac{\dd\rho^2}{4\rho^2}+\bar{g}_{ij}(\rho,x)\dd x^i \dd x^j \,, 
\end{equation}
where the boundary is located at $\rho=0$ and is parametrized by the coordinates $x^i$ with \mbox{$i=0, \ldots, 3$} and $L$ denotes the Anti-de Sitter length scale which we set to unity.
Near the boundary the metric and the scalar field take the form
\begin{eqnarray}\label{seriesVac}
\bar{g}_{ij}(\rho,x)&=&\frac{1}{\rho} \Big[\gamma_{ij}(x) + \rho\, \gamma_{(2)ij}(x)+\rho^2 \gamma_{(4)ij}(x)\nonumber\\
&+&\rho^2\log \rho \,h_{(4)ij}(x)+O(\rho^3) \Big]\,,\\
\Phi(\rho,x)&=&\rho^{1/2}\Big[ \Phi_{(0)}(x)+ \rho \, \Phi_{(2)}(x)\nonumber\\
&+&\rho \,\log \rho\,\psi_{(2)}(x)+O(\rho^2) \Big]\,.   
\end{eqnarray}
The first term $\gamma_{ij}(x)$ in the expansion of the metric is the boundary metric and the term $ \Phi_{(0)}(x)$ plays the role of the inflaton field in the boundary theory.
The Klein--Gordon equation for the scalar field fixes the logarithmic coefficient $\psi_{(2)}$ in terms of $\gamma_{ij}$ and $\Phi_{(0)}$ as
\begin{equation}
\label{psi2}
\psi_{(2)}=\frac{1}{4} \left(\nabla^2\Phi_{(0)}-\frac{1}{6}\Phi_{(0)} R\right)\,. 
\end{equation}
At leading order Einstein's equations determine 
\begin{equation}
\gamma_{(2)ij}=-\frac{1}{2}\left(R_{ij}-\frac{1}{6} R \, \gamma_{ij} \right)
-\frac{\Phi_{(0)}^2}{3} \gamma_{ij}\,.
\end{equation}
The logarithmic part at sub-leading order fixes
\begin{eqnarray}
h_{(4)ij}&=&h_{(4)ij}^{\mbox{\tiny{grav}}}-\frac{1}{12}R_{ij}\Phi_{(0)}^2-\frac{1}{3}\nabla_i\Phi_{(0)}\nabla_j\Phi_{(0)}\nonumber\\
&+&\frac{1}{12}\nabla_i\Phi_{(0)}\nabla^i\Phi_{(0)}\,\gamma_{ij}+\frac{1}{6}\Phi_{(0)}\nabla_i\nabla_j\Phi_{(0)}\nonumber\\
&+&\frac{1}{12}\Phi_{(0)}\,\square_{\gamma}\,\Phi_{(0)}\, \gamma_{ij}\,,
\end{eqnarray}
where the pure gravitational part is given by
\begin{eqnarray}
h_{(4)ij}^{\mbox{\tiny{grav}}}&=&\frac{1}{8}R_{ikjl}R^{kl}-\frac{1}{48}\nabla_i\nabla_j R+\frac{1}{16}\nabla^2R_{ij}-\frac{1}{24}RR_{ij}
\nonumber\\
&+&\left(\frac{1}{96}R^2-\frac{1}{96}\nabla^2R-\frac{1}{32}R_{kl}R^{kl}\right)\gamma_{ij}\,.
\end{eqnarray}
The expectation values of the holographic energy momentum tensor and the scalar field operator follow from variations of the renormalized action of the holographic model
\begin{equation}\label{eq:SholApp}
S_{\rm hol}=S_{\rm bulk}+S_{\rm GHY}+S_{\rm ct}\,.
\end{equation}
The bare bulk action $S_{\rm bulk}$ is defined in Eq.~(5) of the main text and the second term is the Gibbons--Hawking--York (GHY) boundary term
\begin{equation}\label{eq:GHY}
S_{\rm GHY}=\frac{1}{\kappa_5}\int\dd^4x\sqrt{-\gamma}K\,,
\end{equation}
where $K=\gamma^{ij}K_{ij}=\gamma^{ij}\nabla_i n_j$ denotes the trace of the extrinsic curvature of a four-dimensional slice of the bulk geometry near the boundary.
The last contribution in~\eqref{eq:SholApp} is a counter term that is defined on a constant-$\rho$ hypersurface near the boundary, which is necessary to render the on-shell action $S_{\rm hol}$ finite in the limit $\rho\to0$
\begin{eqnarray}\label{eq:Sct}
 S_{\mathrm{ct}}&=& \frac{1}{\kappa_5} \int \dd^4x\sqrt{-\gamma}\Bigg[ \left(-\frac{1}{8}R-\frac{3}{2}-\frac{1}{2}\Phi_{(0)}^2 \right)\nonumber\\ 
&+& \frac{1}{2} \left( \log \rho\right)  \mathcal{A}+\left( \alpha \mathcal{A} + \beta\Phi_{(0)}^4 \right)\Bigg]\,,
\end{eqnarray}
where the constants $\alpha$ and $\beta$ parametrize the residual renomalization-scheme ambiguity of the model.
The holographic conformal anomaly~\cite{Henningson:1998gx,Papadimitriou:2011qb} $\mathcal{A}=\mathcal{A}_g +\mathcal{A}_\phi$ consists of a gravitational part due to the curved boundary geometry
\begin{equation}\label{eq:Ag}
 \mathcal{A}_g = \frac{1}{16}(R^{ij}R_{ij}-\frac{1}{3}R^2)\,,
\end{equation}
and a part due to scalar matter
\begin{equation}\label{eq:Aphi}
 \mathcal{A}_{\Phi_{(0)}}=-\frac{1}{2}\left( \partial_i\Phi_{(0)}\partial^i\Phi_{(0)}+\frac{1}{6}R\Phi_{(0)}^2\right)\,.
\end{equation}

All this together results in the following expression for the holographic stress tensor 
\begin{eqnarray}\label{eq:EMT}
\langle  T_{ij}^{\rm QFT} \rangle&=&\frac{2}{\sqrt{-\gamma}}\frac{\delta S_{\rm hol}}{\delta \gamma^{ij}}\nonumber\\
&=& \frac{2}{\kappa_5} 
\Bigg\{\gamma_{(4)ij}+\frac{1}{8}\left[\mathrm{Tr}\gamma_{(2)}^2-(\mathrm{Tr}\gamma_{(2)})^2\right]\gamma_{ij}\nonumber\\
&-&\frac{1}{2}\gamma_{(2)}^2+\frac{1}{4}\gamma_{(2)ij}\mathrm{Tr}\gamma_{(2)}+\frac{1}{2}\partial_i\Phi_{(0)}\partial_j\Phi_{(0)}\nonumber\\
&+&\left(\Phi_{(0)}\Phi_{(2)}-\frac{1}{2}\Phi_{(0)}\psi_{(2)}-\frac{1}{4}\partial_k\Phi_{(0)}\partial^k\Phi_{(0)}\right)\gamma_{ij}\nonumber\\
&+&\alpha\left(\mathcal{T}^{\gamma}_{ij}+\mathcal{T}_{ij}^\phi\right)+\left(\frac{1}{18}+\beta \right)\Phi_{(0)}^4\gamma_{ij}\Bigg\}\,. 
\end{eqnarray}
The anomalous contributions to the stress tensor are given by
\begin{eqnarray}
\mathcal{T}^{g}_{ij}&=&2h_{(4)ij}\,,\\
\mathcal{T}_{ij}^\phi&=&-\frac{1}{2}\Phi_{(0)}^2R_{ij}-\frac{2}{3}\nabla_i\Phi_{(0)}\nabla_j\Phi_{(0)}+\frac{1}{6}\gamma^{kl}\nabla_k\nabla_l\Phi_{(0)}\gamma_{ij}\nonumber\\
&+&\frac{1}{3}\Phi_{(0)}\nabla_i\nabla_j\Phi_{(0)}+\frac{1}{6}\Phi_{(0)}\square \Phi_{(0)} \gamma_{ij}\nonumber\\
&-&\frac{1}{2}\gamma_{ij}\left(\Phi_{(0)}\square\Phi_{(0)}-\frac{1}{6}R\Phi_{(0)}^2 \right)\,.
\end{eqnarray}
The expectation value of the scalar operator in the field theory is then given by
\begin{eqnarray}\label{VEV}
\langle\mathcal{O}\rangle&=&\frac{1}{\sqrt{-\gamma}}\frac{\delta S_{\rm hol}}{\delta \Phi_{(0)}}\nonumber\\
&=&\frac{2}{\kappa_5}
\left[(1-4\alpha)\psi_{(2)}-2\Phi_{(2)}-4\beta\Phi_{(0)}^3\right]\,.
\end{eqnarray}
The holographic stress tensor satisfies anomaly-corrected Ward identities
\begin{eqnarray}
\nabla^i\langle T_{ij}^{\rm QFT}\rangle&=&-\langle\mathcal{O}\rangle\nabla_j \Phi_{(0)}\,,\label{ex:Ward}\\
\gamma^{ij}\langle T_{ij}^{\rm QFT}\rangle&=&- \Phi_{(0)}\langle\mathcal{O}\rangle+
\frac{1}{\kappa_5} \left(\mathcal{A}_{g}+\mathcal{A}_{\Phi_{(0)}}\right)\,.
\end{eqnarray}
The inflaton field in the main text $\phi=\Phi_{(0)}/\lambda$ is related to the source of the scalar operator $\Phi_{(0)}$ via the coupling constant $\lambda$ and all QFT expectation values are given in units of the bulk gravitational coupling
\begin{eqnarray}
\mathcal{T}_{ij}^{\rm QFT}&=&\kappa_5\langle T^{\rm QFT}_{ij}\rangle\nonumber\\
&=&{\rm diag}\left(\mathcal{E}_{\rm QFT},\mathcal{P}_{\rm QFT},\mathcal{P}_{\rm QFT},\mathcal{P}_{\rm QFT}\right)\,,\\
\mathcal{O}_{\rm QFT}&=&\kappa_5\langle \mathcal{O}\rangle\,.
\end{eqnarray}
Finally, in our setup with dynamical boundary equations, $\alpha$ and $\beta$ renormalize the bare gravitational coupling and the cosmological constant in the boundary theory~\cite{Ecker:2021cvz}
\begin{eqnarray}
\frac{1}{\kappa_4}&=&\frac{1}{\kappa_{4,\rm bare}}+\frac{\alpha}{96\,\kappa_5}\,,\\
\frac{\Lambda_4}{\kappa_4}&=&\frac{\Lambda_{4,\rm bare}}{\kappa_4}-\frac{\beta}{1024\,\pi}\,.
\end{eqnarray} 
We fix $\alpha=0$ and $\beta=\frac{1}{16}$,  because this choice leads to a supersymmetric renormalisation scheme in which the full boundary stress tensor vanishes identically if the boundary metric is flat. 

\subsection{Properties of the holographic QFT}\label{app:hQFT}
\begin{figure}
\center
\includegraphics[width=0.9\linewidth]{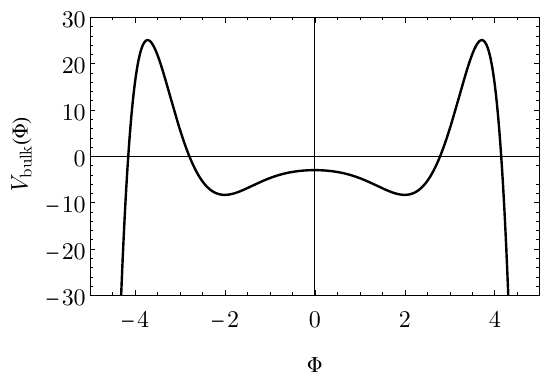}
\caption{
Scalar field potential of the holographic model.
}
\label{fig:Vbulk}
\end{figure}

Here we review some basic properties of the holographic QFT used in this work (see also~\cite{Casalderrey-Solana:2020vls}).
This theory has a relevant scalar operator that is dual to a bulk scalar field $\Phi$.
For convenience, we repeat here the corresponding Einstein-dilaton type bulk action 
\begin{equation}\label{S1}
 S_{\rm bulk}=\frac{2}{\kappa_5}\int \dd^5x\sqrt{-g}\left(\frac{1}{4}\mathcal{R}-\frac{1}{2}(\partial\Phi)^2-V_{\rm bulk}(\Phi)\right)\,,
\end{equation}
where $\kappa_5$ denotes the bulk gravitational coupling, $\mathcal{R}$ is the Ricci scalar associated to the bulk metric $g_{\mu\nu}$ and $\Phi$ is the bulk scalar field with potential
\begin{equation}\label{W1}
V_{\rm bulk}(\Phi)=\frac{1}{L^2}\left(-3-\frac{3 \Phi ^2}{2}-\frac{\Phi ^4}{3}+\frac{11 \Phi ^6}{96}-\frac{\Phi ^8}{192}\right)\,.
 \end{equation}
The bulk potential $V_{\rm bulk}(\Phi)$ has several extrema and is shown in Fig.~\ref{fig:Vbulk}. 
The most interesting for us is the maximum at $\Phi=0$, which corresponds to an UV fixed point, or a CFT$_{\rm UV}$.
On the two sides of this maximum are two symmetric minima at $\Phi=\pm 2$ that correspond to two copies of an IR conformal theory CFT$_{\rm IR}$.
As the potential has a $\Phi \to -\Phi$ symmetry, the two minima correspond to the same CFT$_{\rm IR}$.
Around the maximum, the dimension of the relevant scalar operator is $\Delta_{\rm UV}=3$, while at the minima the dimension of the same scalar operator is $\Delta_{\rm IR}=25/6$.
The relevant coupling in the CFT$_{\rm UV}$
has mass dimension one, and is therefore like a fermion mass scale.
Our QFT is a RG flow between CFT$_{\rm UV}$ and CFT$_{\rm IR}$ that is driven by the source of the scalar operator with mass scale $m$.
Moreover, the QFT has the symmetry $m\to -m$.
This QFT is therefore massless in the IR, with the massless (and strongly coupled) degrees of freedom being those of CFT$_{IR}$.
Moreover, although the QFT is strongly coupled at all scales, it is a non-confining QFT.

\begin{figure*}[htb]
\center
\includegraphics[width=0.94\linewidth]{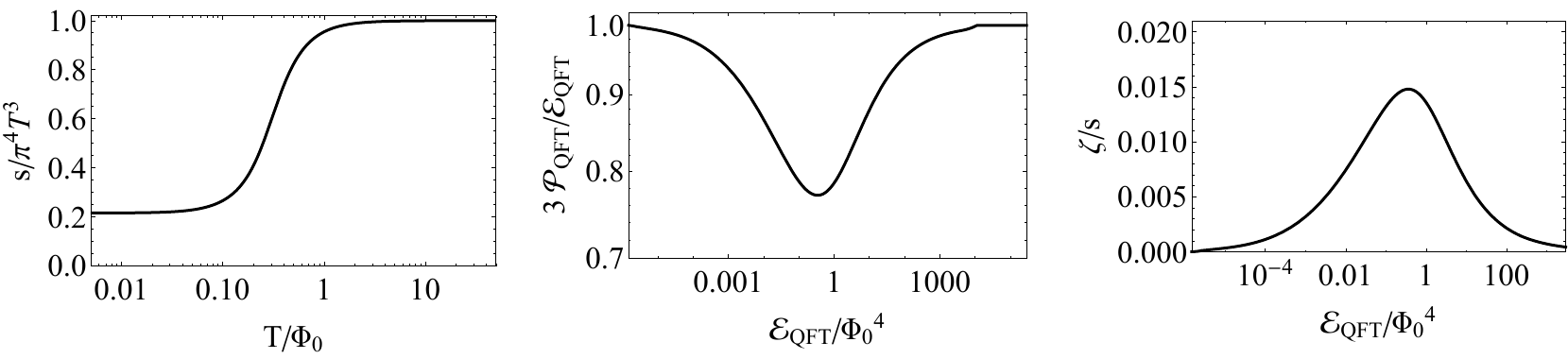}
\caption{We show the thermodynamical entropy density (left), pressure (middle) and bulk viscosity (right) as a function of temperature or energy density. 
This figure is reproduced from~\cite{Casalderrey-Solana:2020vls} (see also \cite{Attems:2016ugt}).
}
\label{fig:thermodynamics}
\end{figure*}

The thermodynamic and transport properties of the model were analyzed in~\cite{Attems:2016ugt}.
At $T>0$ the theory is entering the black-hole phase 
and remains in it, all the way to $T\to\infty$.
In Fig.~\ref{fig:thermodynamics} (reproduced from~\cite{Casalderrey-Solana:2020vls}) we show from left to right the dimensionless entropy ratio $s/T^3$ as a function of the (dimensionless) temperature, the ratio $3{\cal P}/{\cal E}$ of the pressure to the energy density and 
ratio of the bulk viscosity to the entropy density, $\zeta/s$, as function of energy density.
All these quantities asymptote to their conformal values at small and large temperatures or energy densities.

Finally, we comment here on the evolution of the QFT when it is coupled to the inflaton. 
The inflaton field is by construction proportional to the mass scale of the QFT.
When the inflaton slow-rolls in the potential on the left side of Fig.~2 in the main text, the mass scale of the QFT starts at large negative values and slowly increases towards zero.
This represents an inverse RG flow that is driven by the cosmological evolution of the inflaton.
Once the inflaton settles at the minimum of the potentials at $\Phi=0$, the mass scale becomes zero and leaves the QFT at its UV limit, namely CFT$_{UV}$.

\subsection{Solving the Bulk Model Numerically}\label{app:IVP}

The action in Eq.~(1) of the main text results in a coupled set of equations of motion for the bulk that are the five-dimensional Einstein--Klein--Gordon equations
\begin{eqnarray}
\mathcal{R}_{\mu\nu}-\frac{1}{2}\mathcal{R}\,g_{\mu\nu}&=&2\partial_\mu\Phi\partial_\nu\Phi-g_{\mu\nu}\left(2V_{\rm bulk}+(\partial\Phi)^2\right)\,,\\
\square_g\Phi&=&\frac{\partial V_{\rm bulk}}{\partial \Phi}\,.
\end{eqnarray}
The method we use to solve the corresponding initial value problem for fixed boundary conditions was first presented in \cite{Chesler:2008hg} for pure gravity and further reviewed in \cite{Chesler:2013lia, vanderSchee:2014qwa}. 
For the case of dynamical boundary conditions as studied here, the method was first extended numerically in \cite{Ecker:2021cvz}.
Here we give a summary of this method.

For the numerical treatment of the initial value problem it is convenient to use generalized Eddington--Finkelstein (EF) coordinates rather than FG gauge to parametrize the bulk geometry and the scalar field
\begin{eqnarray}\label{metricEF}
\dd s_{\rm bulk}^2&=&g_{\mu\nu}\dd x^\mu\dd x^\nu\\
   &=&-A(r,t)\dd t^2+2\dd r\dd t+S(r,t)^2\dd\vec{x}^2\,,\\
\Phi&=&\Phi(r,t)\,, 
\end{eqnarray}
where the asymptotic boundary is located at $r= \infty$.
In this gauge the coupled set of Einstein and scalar field equations result in the following nested set of ODEs
\begin{eqnarray}\label{eq:CharEOM}
 S''&=&-\frac{2}{3} S \left(\Phi '\right)^2 \,,\label{eq:Spp}\\
 \dot{S}'&=&-\frac{2 \dot{S} S'}{S}-\frac{2 S V}{3}\,, \label{eq:Sd}\\
 \dot{\Phi }'&=&\frac{V'}{2}-\frac{3\dot{S} \Phi '}{2S}-\frac{3S'\dot{\Phi }}{2S}\,, \label{eq:phid} \\
 A''&=&\frac{12 \dot{S} S'}{S^2}+\frac{4 V}{3}-4 \dot{\Phi } \Phi'\,,\label{eq:App} \\
 \ddot{S}&=&\frac{\dot{S} A'}{2}-\frac{2 S \dot{\Phi }^2}{3} \label{eq:Sdd}\,,
\end{eqnarray}
where a prime denotes a radial derivative, $f'\equiv \partial_r f$, and an overdot is short-hand for the modified derivative $\dot{f}\equiv \partial_t f +\frac{1}{2}A \partial_r f $. The beauty of the scheme of this so-called characteristic formulation is that specifying $\Phi(z)$ leads to $\partial_t \Phi(z)$ through this nested set of ODEs, which is much simpler than the typical PDE system encountered in numerical relativity.
We solve the initial value problem using the procedure explained in \cite{Casalderrey-Solana:2020vls} imposing the Friedmann--Lemaître--Robertson--Walker metric
\begin{equation}
 \dd s^2=\gamma_{ij}\dd x^i \dd x^j=- \dd t^2 + a(t)^2 \dd\vec{x}^2\,,
\end{equation}
which leads to the boundary condition $S(r) = r\,a(t)+\mathcal{O}(r^0)$ for Eq.~\eqref{eq:Spp}.
The boundary conditions for Eq.~\eqref{eq:Sd} and Eq.~\eqref{eq:phid} is fixed by demanding regularity near the boundary, while for Eq.~\ref{eq:App} they are fixed by the energy density $\mathcal{E}$.
The energy density is evolved by using conservation of the stress-energy tensor or alternatively by using Eq.~\eqref{eq:Sdd}.

In practice it is difficult to solve these equations directly.
For numerical efficiency it is better to switch coordinates to $z=1/r$ and to perform a near-boundary (NB) expansion of both the equations and the functions $S$, $A$ and $\Phi$.
One then redefines $\Phi(r) \equiv \Phi_{NB}(r)+r^{-3}\tilde{\Phi}(r)$ with $\Phi_{NB}(r)$ containing near-boundary terms up to $\mathcal{O}(r^{-2})$ and $\mathcal{O}(r^{-4} \log(r))$ (and analogously for $S$ and $A$).
When using spectral methods, it is especially important to subtract a high number of logarithmic NB terms, as spectral methods rely on regularity of the functions presented.
Lastly, it is convenient to apply a gauge transformation $r\rightarrow r+\xi(t)$ such that the apparent horizon (AH) remains at a constant value of the $r$ coordinate.
Since the condition for the location of the AH equals $\dot{S}=0$ the equation for $\xi(t)$ can be obtained by solving $\partial_t\dot{S}=0$ on the AH (note that in Fig.~5 of the main text we do not apply this gauge transformation for clarity).

The methods to couple the equations with our boundary Friedmann+inflaton equations are exactly the same as in~\cite{Ecker:2021cvz} with the only exception that we now have a dynamical source for the scalar field as is detailed in the main text.
For completeness we note that we use a pseudospectral grid with 5 domains each having 7 grid points. In the simulations we use $\kappa_5=1$ and rescale to the chosen $\kappa_5$ only when plotting results.
We use timesteps of $\delta t=0.0005$ and filter then numerical functions every $50$ timesteps by interpolating back and forth to a spectral grid with 5 points (see also \cite{Chesler:2013lia}).
We start with a radial gauge transformation $r\rightarrow r+\xi$ with $\xi=1.709$.
This fixes the apparent horizon at $z=1/r= 0.36$ and the evolution equations for $\xi(t)$ guarantee that the horizon stays there (nevertheless, every 100 timesteps we perform a tiny gauge transformation to bring it back to $z=0.36$). The full evolution then takes around 200 hours on a single core using \emph{Mathematica} 11.

For reference we note that the temperature as derived from the surface gravity $\kappa$ can be obtained from $T_{\rm AH} = \kappa/2\pi = -\frac{z^2}{4\pi} \partial_z A$ evaluated at the location of the apparent horizon. 
For the event horizon this requires solving the differential equation $\partial_t r_{\rm EH}=-\tfrac{1}{2} A$ with boundary condition $A(t, r_{\rm EH}(t=\infty)) = 0$. 
This reflects the fact that the event horizon is teleological, e.g. it depends on the future spacetime.
In our simulations we go to a finite time where the geometry is sufficiently constant such that a full solution of the event horizon can be obtained.

The full code can be downloaded at \href{http://wilkevanderschee.nl}{wilkevanderschee.nl}.

\end{document}